\documentclass{article}
\usepackage{amsmath}

\input{tcilatex}

\begin{document}

\title{The Role of Measurement in Quantum Games}
\author{ Ahmad Nawaz\thanks{%
ahmad@ele.qau.edu.pk} and A. H. Toor\thanks{%
ahtoor@qau.edu.pk} \\
%EndAName
Department of Physics, Quaid-i-Azam University, \ \ \ \ \ \\
Islamabad 45320, Pakistan.}
\maketitle

\begin{abstract}
The game of Prisoner Dilemma is analyzed to study the role of measurement
basis in quantum games. Four different types of payoffs for quantum games
are identified on the basis of different combinations of initial state and
measurement basis. A relation\ among these different payoffs is established.
\end{abstract}

\section{Introduction}

Game theory deals with a situation in which two or more parties compete to
maximize their respective payoffs by playing suitable strategies according
to the known payoff matrix. Extension of game theory to quantum domain with
quantization of the strategy space has shown clear advantage over classical
strategies \cite{meyer,eisert,marinatto,nawaz1}. A detailed description on
classical and quantum game theory can be found in \cite{neumann,lee}

In quantum version of the game arbiter prepares an initial quantum state and
passes it on to the players (generally referred as Alice and Bob). After
applying their local operators (or strategies) the players return the state
to arbiter who then announces the payoffs by\ performing a measurement with
the application of suitable payoff operators depending on the payoff matrix
of the game. The role of the initial quantum state remained an interesting
issue in quantum games \cite{eisert,marinatto,azhar,nawaz1}. However, the
importance of the payoff operators used by arbiter to perform measurement to
determine the payoffs of the players remained unnoticed. In our earlier
paper \cite{nawaz} we have pointed out the importance of measurement basis
in quantum games. It was shown that if the arbiter is allowed to perform the
measurement in the entangled basis interesting situations could arise which
were not possible in the frame work of Eisert et. al. \cite{eisert} and
Marinatto et. al. \cite{marinatto} schemes. In this paper we further extend
our earlier work to investigate the role of measurement basis in quantum
games by taking Prisoner Dilemma as an example. In this scenario quantum
payoffs are divided into four different categories on the basis of initial
state and measurement basis. These different situations arise due the
possibility of having product or entangled initial state and then applying
product or entangled basis for the measurement \cite{pati,xyz}. In the
context of our generalized framework for quantum games, the four different
types of payoffs are\emph{\ }

(i) $\$_{PP}$\ is the payoff when the initial quantum state is of the
product form and product basis are used for measurement to determine the
payoff.

(ii) $\$_{PE}$\ is the payoff when the initial quantum state is of the
product form and entangled basis are used for measurement to determine the
payoff.

(iii) $\$_{EP}$\ is the payoff when the initial quantum state is entangled
and product basis are used for measurement to determine the payoff.

(iv) $\$_{EE}$\ is the payoff when the initial quantum state is entangled
and entangled basis are used for measurement to determine the payoff.

Our results show that these payoffs obey a relation, $\$_{PP}<\$_{PE}=%
\$_{EP}<\$_{EE}$ at the Nash Equilibrium (NE).\ This is also\ interesting to
note that the role of entangled and /or product input and \ entangled \
and/or product measurement in this relation\ is\ very similar to its role in
the existing relation for the classical capacities of the quantum channels.
It is shown in the Ref. \cite{king} that for a quantum channel the
capability to transmit maximum classical information, called the classical
channel capacity $C$\ of a quantum channel, a relation of the form $%
C_{PP}<C_{PE}=C_{EP}<C_{EE}$\ \ holds. In this paper we have not tried to
investigate the possible relationship between channel capacity and payoff's.

\section{\label{prisoner}Prisoner Dilemma}

In the game of Prisoner Dilemma two prisoners are being interrogated in
separate cells for a crime they have committed together. The two possible
moves for these prisoners are to cooperate ($C$) or to defect ($D$). They
are not allowed to communicate but have access to the following payoff
matrix:

\begin{equation}
\text{Alice}%
\begin{array}{c}
C \\ 
D%
\end{array}%
\overset{\text{Bob}}{\overset{%
\begin{array}{cc}
C\text{ \ \ \ \ } & D%
\end{array}%
}{\left[ 
\begin{array}{cc}
\left( 3,3\right) & \left( 0,5\right) \\ 
\left( 5,0\right) & \left( 1,1\right)%
\end{array}%
\right] }},  \label{classical payoff}
\end{equation}%
\ It can be seen from the Eq. (\ref{classical payoff}) that $D$ is the
dominant strategy for the two players. Therefore, rational reasoning forces
each player to play $D$ causing \ ($D,D$) as the Nash equilibrium of the
game with payoffs \ (1,1), i.e., 1 for both. The players could have got
higher payoffs had both of them decided to play $C$ instead of $D$. This is
the dilemma in this game \cite{flood}.\emph{\ }Eisert et. al \cite{eisert}
analyzed this game in quantum domain and showed that there exist a suitable
quantum strategy for which the dilemma is resolved. They also pointed out a
quantum strategy which always wins over all classical strategies.

In our generalized version of quantum games the arbiter prepares the initial
state of the form

\begin{equation}
\left| \psi _{in}\right\rangle =\cos \frac{\gamma }{2}\left| CC\right\rangle
+i\sin \frac{\gamma }{2}\left| DD\right\rangle .  \label{state in}
\end{equation}%
Here $\left| C\right\rangle $ and $\left| D\right\rangle $, represent
vectors in the strategy space corresponding to Cooperate and Defect,
respectively\emph{\ }with\emph{\ }$\gamma \in \left[ 0,\pi \right] $\textbf{%
\ }\emph{. }The strategy of each of the\ players can be represented by the
unitary operator $U_{i}$\ of the form 
\begin{equation}
U_{i}=\cos \frac{\theta _{i}}{2}R_{i}+\sin \frac{\theta _{i}}{2}P_{i},\text{
\ \ \ }  \label{combination}
\end{equation}%
where $i=1$\ or $2$\ and $R_{i}$, $P_{i}$\emph{\ }are the unitary operators
defined as:

\begin{align}
R_{i}\left| C\right\rangle & =e^{i\phi _{i}}\left| C\right\rangle ,\text{ \
\ }R_{i}\left| D\right\rangle =e^{-i\phi _{i}}\left| D\right\rangle ,  \notag
\\
P_{i}\left| C\right\rangle & =-\left| D\right\rangle ,\text{ \ \ \ \ \ }%
P_{i}\left| D\right\rangle =\left| C\right\rangle .  \label{oper}
\end{align}%
\textbf{\ }Here we restrict our treatment to two parameter set of strategies
($\theta _{i},\phi _{i}$) for mathematical simplicity in accordance with the
Ref. \cite{eisert}. After the application of the strategies, the initial
state given by the eq. (\ref{state in}) transforms to 
\begin{equation}
\left| \psi _{f}\right\rangle =(U_{1}\otimes U_{2})\left| \psi
_{in}\right\rangle .  \label{final}
\end{equation}%
and using Eqs. (\ref{oper}) and (\ref{final}) the above expression becomes 
\begin{align}
\left| \psi _{f}\right\rangle & =\cos \left( \gamma /2\right) \left[ \cos
\left( \theta _{1}/2\right) \cos \left( \theta _{2}/2\right) e^{i\left( \phi
_{1}+\phi _{2}\right) }\left| CC\right\rangle -\cos \left( \theta
_{1}/2\right) \sin \left( \theta _{2}/2\right) e^{i\phi _{1}}\left|
CD\right\rangle \right.  \notag \\
& -\left. \cos \left( \theta _{2}/2\right) \sin \left( \theta _{1}/2\right)
e^{i\phi _{2}}\left| DC\right\rangle +\sin \left( \theta _{1}/2\right) \sin
\left( \theta _{2}/2\right) \left| DD\right\rangle \right]  \notag \\
& +i\sin \left( \gamma /2\right) \left[ \cos \left( \theta _{1}/2\right)
\cos \left( \theta _{2}/2\right) e^{-i(\phi _{1}+\phi _{2})}\left|
DD\right\rangle +\cos \left( \theta _{1}/2\right) \sin \left( \theta
_{2}/2\right) e^{-i\phi _{1}}\left| DC\right\rangle \right.  \notag \\
& +\left. \cos \left( \theta _{2}/2\right) \sin \left( \theta _{1}/2\right)
e^{-i\phi _{2}}\left| CD\right\rangle +\sin \left( \theta _{1}/2\right) \sin
\left( \theta _{2}/2\right) \left| CC\right\rangle \right] .
\label{state fin}
\end{align}%
The operators used by the arbiter to determine the payoff for Alice and Bob
are

\begin{align}
P_{A}& =3P_{CC}+P_{DD}+5P_{DC}  \notag \\
P_{B}& =3P_{CC}+P_{DD}+5P_{CD}  \label{pay-operator}
\end{align}%
where 
\begin{subequations}
\label{oper a}
\begin{align}
P_{CC}& =\left| \psi _{CC}\right\rangle \left\langle \psi _{CC}\right| \text{%
, \ }\left| \psi _{CC}\right\rangle =\cos \left( \delta /2\right) \left|
CC\right\rangle +i\sin \left( \delta /2\right) \left| DD\right\rangle ,
\label{oper 1} \\
P_{DD}& =\left| \psi _{DD}\right\rangle \left\langle \psi _{DD}\right| \text{%
, \ }\left| \psi _{DD}\right\rangle =\cos \left( \delta /2\right) \left|
DD\right\rangle +i\sin \left( \delta /2\right) \left| CC\right\rangle ,
\label{oper 2} \\
P_{DC}& =\left| \psi _{DC}\right\rangle \left\langle \psi _{DC}\right| \text{%
, \ }\left| \psi _{DC}\right\rangle =\cos \left( \delta /2\right) \left|
DC\right\rangle -i\sin \left( \delta /2\right) \left| CD\right\rangle ,
\label{oper 3} \\
P_{CD}& =\left| \psi _{CD}\right\rangle \left\langle \psi _{CD}\right| \text{%
, \ }\left| \psi _{CD}\right\rangle =\cos \left( \delta /2\right) \left|
CD\right\rangle -i\sin \left( \delta /2\right) \left| DC\right\rangle ,
\label{oper 4}
\end{align}%
with\emph{\ }$\delta \in \left[ 0,\pi \right] $. Above payoff operators
reduce to that of Eisert's scheme for $\delta $ equal to $\gamma ,$ which
represents the entanglement of the initial state \cite{eisert}. And for $%
\delta =0$ above operators transform into that of Marinatto and Weber's
scheme \cite{marinatto}. In our generalized quantization scheme, payoffs for
the players are calculated as 
\end{subequations}
\begin{eqnarray}
\$^{A}(\theta _{1},\phi _{1},\theta _{2},\phi _{2}) &=&\text{Tr}(P_{A}\rho
_{f})\text{,}  \notag \\
\$^{B}(\theta _{1},\phi _{1},\theta _{2},\phi _{2}) &=&\text{Tr}(P_{B}\rho
_{f}),  \label{payoff}
\end{eqnarray}%
\emph{\ } where $\rho _{f}=\left| \psi _{f}\right\rangle \left\langle \psi
_{f}\right| $ is the density matrix for the quantum state given by (\ref%
{state fin}) and Tr represents the trace of a\emph{\ }matrix. Using Eqs. (%
\ref{state fin}), (\ref{oper a}), and (\ref{payoff}), we get the following
payoffs 
\begin{eqnarray}
\$^{A}\left( \theta _{i},\phi _{j}\right)  &=&\sin ^{2}\left( \theta
_{1}/2\right) \sin ^{2}\left( \theta _{2}/2\right) \left[ \cos ^{2}\left( 
\frac{\gamma +\delta }{2}\right) +3\sin ^{2}\left( \frac{\gamma -\delta }{2}%
\right) \right]   \notag \\
&&+\cos ^{2}\left( \theta _{1}/2\right) \cos ^{2}\left( \theta _{2}/2\right) 
\left[ 2+\cos \gamma \cos \delta +2\cos \left( 2\delta \left( \phi _{1}+\phi
_{2}\right) \right) \sin \gamma \sin \delta \right]   \notag \\
&&-\sin \theta _{1}\sin \theta _{2}\sin \left( \phi _{1}+\phi _{2}\right) 
\left[ \sin \gamma -\sin \delta \right] +\frac{5}{4}\left[ 1-\cos \theta
_{1}\cos \theta _{2}\right]   \notag \\
&&+\frac{5}{4}\left( \cos \theta _{2}-\cos \theta _{1}\right) \left[ \cos
\gamma \cos \delta +\cos \left( 2\phi _{1}\right) \sin \gamma \sin \delta %
\right] .  \label{payoff-general-a}
\end{eqnarray}%
The payoff of player $B$ can be found by interchanging $\theta
_{1}\longleftrightarrow $\ $\theta _{2}$\ and $\phi _{1}\longleftrightarrow
\phi _{2}$ in the Eq. (\ref{payoff-general-a})$.$ There can be four types of
payoffs for each player\ for different combinations of $\delta $ and $\gamma 
$. In the following $\$_{PP}\left( \theta _{1},\theta _{2}\right) $ means
payoffs of the players when the initial state of the game is product state
and payoff operator used by arbiter for measurement is also in the product
form $(\gamma =0,\delta =0)$ and $\$_{EP}\left( \theta _{1},\theta _{2},\phi
_{1},\phi _{2}\right) $ means the payoffs for entangled input state when the
payoff operator used for measurement is in the product form, i.e., $(\gamma
\neq 0,\delta =0)$. Similarly $\$_{PE}\left( \theta _{1},\theta _{2},\phi
_{1},\phi _{2}\right) $ and $\$_{EE}\left( \theta _{1},\theta _{2},\phi
_{1},\phi _{2}\right) $ can also be interpreted. Therefore, for different
values of $\delta $ and $\gamma $ the\emph{\ }following four cases can be
identified:

\textbf{Case (a) }When\textbf{\ }$\delta $\textbf{\ }$=$\textbf{\ }$\gamma
=0,$ the Eq.(\ref{payoff-general-a}), becomes 
\begin{subequations}
\label{SPP-prisoner}
\begin{equation}
\$_{PP}^{A}\left( \theta _{1},\theta _{2}\right) =3\cos ^{2}\left( \theta
_{1}/2\right) \cos ^{2}\left( \theta _{2}/2\right) +\sin ^{2}\left( \theta
_{1}/2\right) \sin ^{2}\left( \theta _{2}/2\right) +5\sin ^{2}\left( \theta
_{1}/2\right) \cos ^{2}\left( \theta _{2}/2\right)  \label{SPP-prisoner-a}
\end{equation}%
This situation corresponds to the classical game where each player play,\ $%
C, $ with probability $\cos ^{2}\left( \theta _{i}/2\right) $ with $i=1,2$ %
\cite{eisert1}$.$ The Nash equilibrium corresponds to $\theta _{1}=\theta
_{2}=\pi ,$ i.e., $(D,D)$ with payoffs for both the players as

\end{subequations}
\begin{equation}
\$_{PP}^{A}(\theta _{1}=\pi ,\theta _{2}=\pi )=\$_{PP}^{B}(\theta _{1}=\pi
,\theta _{2}=\pi )=1.  \label{SPP-Nash}
\end{equation}

\textbf{Case (b) }When $\gamma =0,\delta $\textbf{\ }$\neq 0,$ in the Eq. (%
\ref{payoff-general-a}), then the game has two Nash equilibria one at $%
\theta _{1}=\theta _{2}=0$ when $\sin ^{2}\left( \delta /2\right) \geq \frac{%
2}{3}$ \ and the other at $\theta _{1}=\theta _{2}=\pi $ when $\sin
^{2}\left( \delta /2\right) \leq \frac{1}{3}$. The corresponding payoffs for
these Nash equilibria are

\begin{eqnarray}
\$_{PE}^{A}(\theta _{1} &=&0,\theta _{2}=0)=\$_{PE}^{B}(\theta _{1}=0,\theta
_{2}=0)=3-2\sin ^{2}\left( \delta /2\right) ,  \notag \\
\$_{PE}^{A}(\theta _{1} &=&\pi ,\theta _{2}=\pi )=\$_{PE}^{B}(\theta
_{1}=\pi ,\theta _{2}=\pi )=1+2\sin ^{2}\left( \delta /2\right) .
\label{SPE-nash}
\end{eqnarray}%
Here in this case at NE the payoffs are independent of $\phi _{1},\phi _{2}.$
It is clear that the above payoffs for all the allowed values of $\delta $
remain less than 3, which is the optimal payoff for the two players if they
cooperate.

\textbf{Case (c) }For $\gamma \neq 0,$ and $\delta $\textbf{\ }$=0,$ the
Eqs. (\ref{payoff-general-a}) again gives two Nash equilibria one at $\theta
_{1}=\theta _{2}=0$ when $\sin ^{2}\left( \gamma /2\right) \geq \frac{2}{3}$
\ and the other at $\theta _{1}=\theta _{2}=\pi $ when $\sin ^{2}\left(
\gamma /2\right) \leq \frac{1}{3}$. The corresponding payoffs are

\begin{eqnarray}
\$_{EP}^{A}(\theta _{1} &=&0,\theta _{2}=0)=\$_{EP}^{B}(\theta _{1}=0,\theta
_{2}=0)=3-2\sin ^{2}\left( \gamma /2\right) ,  \notag \\
\$_{EP}^{A}(\theta _{1} &=&\pi ,\theta _{2}=\pi )=\$_{EP}^{B}(\theta
_{1}=\pi ,\theta _{2}=\pi )=1+2\sin ^{2}\left( \gamma /2\right) .
\label{SEP-nash}
\end{eqnarray}%
It can be seen that the payoffs at both Nash equilibrium for allowed values
of $\sin ^{2}\frac{\gamma }{2}$\ remain less than 3. From the Eqs. (\ref%
{SPE-nash}) and (\ref{SEP-nash}), it is also clear that $\$_{EP}^{A}(0,0)=%
\$_{PE}^{A}(\pi ,\pi )$ only for $\delta =\gamma .$

\textbf{Case (d) }When$\ \gamma =\delta $\textbf{\ }$=\pi /2,$ Eqs. (\ref%
{payoff-general-a}) becomes 
\begin{subequations}
\label{SEE-prisoner}
\begin{align}
\$_{EE}^{A}\left( \theta _{1},\theta _{2},\phi _{1},\phi _{2}\right) & =3 
\left[ \cos \left( \theta _{1}/2\right) \cos \left( \theta _{2}/2\right)
\cos \left( \phi _{1}+\phi _{2}\right) \right] ^{2}  \notag \\
& +\left[ \sin \left( \theta _{1}/2\right) \sin \left( \theta _{2}/2\right)
+\cos \left( \theta _{1}/2\right) \cos \left( \theta _{2}/2\right) \sin
\left( \phi _{1}+\phi _{2}\right) \right] ^{2}  \notag \\
& +5\left[ \sin \left( \theta _{1}/2\right) \cos \frac{\theta _{2}}{2}\cos
\phi _{2}-\cos \left( \theta _{1}/2\right) \sin \left( \theta _{2}/2\right)
\sin \phi _{1}\right] ^{2}  \notag \\
&  \label{SEE-prisoner-b}
\end{align}%
This payoff is same\emph{\ }as found by Eisert et. al. \cite{eisert} and $%
\theta _{1}=\theta _{2}=0,\phi _{1}=\phi _{2}=\frac{\pi }{2}$ is the Nash
equilibrium \cite{eisert} of the game that gives the payoffs for both
players as 
\end{subequations}
\begin{equation}
\$_{EE}^{A}(0,0,\frac{\pi }{2},\frac{\pi }{2})=\$_{EE}^{B}(0,0,\frac{\pi }{2}%
,\frac{\pi }{2})=3  \label{SEE-nash}
\end{equation}%
Comparing eqs. (\ref{SPP-Nash},\ref{SPE-nash},\ref{SEP-nash},\ref{SEE-nash}%
), it is evident that 
\begin{equation}
\$_{EE}^{l}(0,0,\frac{\pi }{2},\frac{\pi }{2})>\left( \$_{PE}^{l}(\theta
_{1}=k,\theta _{2}=k),\$_{EP}^{l}(\theta _{1}=k,\theta _{2}=k)\right)
>\$_{PP}^{l}(\theta _{1}=\pi ,\theta _{2}=\pi )
\end{equation}%
and 
\begin{equation}
\$_{PE}^{l}(\theta _{1}=k,\theta _{2}=k)=\$_{EP}^{l}(\theta _{1}=k,\theta
_{2}=k)\text{ for }\gamma =\delta
\end{equation}%
with $k=0,\pi $ and $l=A,B$. This expression shows the crucial role of
entanglement in quantum games. The combination of initial entangled state
with entangled payoff operators gives higher payoffs as copmared to all
other combinations of $\gamma $\ and $\delta $.

\section{Conclusion}

\emph{\ }In quantum games the arbiter (the referee) prepares an initial
quantum state and passes it on to the players (Alice and Bob). After
applying their local operators (their strategies) the players return their
state to the arbiter. The arbiter then performs a measurement on the final
state by applying the payoff operators to determine the payoffs of the
player on the basis of payoff matrix of the game. In our earlier paper \cite%
{nawaz}, we pointed out the importance of measurement in the quantum games.
Here we extended our earlier work,\ by taking Prisoner Dilemma game as an
example and showed that depending on the initial states and type of
measurement (product or entangled), quantum payoffs in games can be
categories in to four different types. These four categories are $%
\$_{PP},\$_{PE},\$_{EP},\$_{EE}$ where $P,$ and $E$\ are abbreviations for
the product and entanglement at input and output. It is shown that there
exists a relation of the form $\$_{PP}<\$_{PE}=\$_{EP}<\$_{EE}$\ among
different payoffs at the NE.

\end{document}